\documentclass[twocolumn,
]{revtex4-1}

\usepackage{graphicx} 
\usepackage[dvipsnames]{xcolor}
\usepackage{braket}
\usepackage{nicefrac, xfrac}
\usepackage[separate-uncertainty=false]{siunitx}

\usepackage{amsmath}
\usepackage{amsfonts}
\usepackage{amssymb}
\usepackage{multirow}

\usepackage[normalem]{ulem}

\begin{document}

\title{Ultracold atomic spin mixtures\\ in ultrastable magnetic field environments}

\author{Riccardo Cominotti$^{1,2}$}  
\author{Chiara Rogora$^{1,2}$} 
\author{Alessandro Zenesini$^{1,2}$}
\author{Giacomo Lamporesi$^{1,2}$}
\author{Gabriele Ferrari$^{1,2}$}

\affiliation{$^1$Pitaevskii BEC Center, CNR-INO and Dipartimento di Fisica, Universit\`a di Trento, 38123 Trento, Italy}

\affiliation{$^2$Trento Institute for Fundamental Physics and Applications, INFN, 38123 Trento, Italy}

\begin{abstract}
Ultracold atomic spin mixtures develop rich and intriguing magnetic properties when an external radiation coherently couples different spin states. In particular, the coupled mixture may acquire a critical behavior when the spin interactions equal the coupling energy. 
However, atomic mixtures generally feature a relatively high sensitivity to magnetic fields that can set a limitation to the observable phenomena.
In this article, we present an overview of experimental studies of magnetism based on superfluid multicomponent gases in an ultrastable magnetic field environment, which recently became available.
\end{abstract}

\maketitle

\section{Introduction}
When two fluids are mixed together, the properties of the mixture can be quite different from those of their individual constituents. For instance, the critical points can dramatically change and mixing-demixing dynamics starts playing a relevant role. 

In recent years, experimental advances in the manipulation of ultracold gases have opened the route for the study of mixtures of quantum fluids. Among the many experimental realizations, an intriguing platform is given by spin mixtures, i.e., homonuclear mixtures realized with atoms occupying different hyperfine states. If all magnetic sublevels of a given hyperfine state are present in the system, such mixtures are usually referred to as spinor gases \cite{StamperKurn2013}, and the wavefunction acquires a multicomponent nature. As a result, the physics of the system is characterized by competition among all the interaction constants, leading to magnetic ordering or fragmentation \cite{Miesner1999, Evrard21}. 

Even a simple two-component spin mixture \cite{Lamporesi2023} allows for a plethora of experiments spanning from magnetism, to defect formation and to cosmology. This is favored by the additional possibility to coherently couple the target atomic states with external resonant radiation \cite{Recati2022}. The presence of a coherent coupling greatly affects the ground state structure and the spectrum of elementary excitations in the system, as it was considered in Ref.~\cite{Abad2013}. 

Unless using special spin mixtures characterized by an energy difference between the atomic states insensitive to the first order to magnetic field variations \cite{Hall1998,Nicklas2011}, the stability against magnetic field fluctuations represents a major challenge in the realization of spin mixtures. In fact, the reproducibility on spin manipulation strongly depends on the control over the energy difference of levels typically featuring different magnetic dipole moments.
The problem becomes more relevant when the coupling radiation is not only used for fast spin manipulation schemes but is applied over long periods. The latter imposes strict constraints on the energy levels' stability to preserve coherence under relatively weak coherent coupling conditions.

In the following sections, we introduce the relevant theoretical background and the experimental platform used in a series of experiments on sodium two-component spin mixtures, investigated in a magnetically shielded environment \cite{Farolfi2019}. 
Then we report the main results of studies on the ground state properties (the spectrum of linear excitations and the para- to ferromagnetism phase transition) and on the out-of-equilibrium dynamics of such a mixture.

\section{Theory}
\label{Theory}

The ground state properties and the dynamics of a bare spin mixture, where two superfluids with atomic densities $n_{\uparrow}$ and $n_{\downarrow}$ coexist, originate from the presence of interactions between atoms in different spin states, which couple the two single-component Gross-Pitaevskii equations. At the mean-field level, the coupling occurs via the intercomponent interaction energy $\propto g_{\uparrow \downarrow} \sqrt{n_{\uparrow}n_{\downarrow}}$, where $g_{\uparrow \downarrow}$ is the intercomponent coupling constant. 
For finite values of $g_{\uparrow \downarrow}$ the mixture excitations can be conveniently described in terms of two independent channels, the total density $n = n_{\uparrow} + n_{\downarrow}$ and the spin channel $s = n_{\uparrow} - n_{\downarrow}$. The two correspond to ``in-phase" and ``out-of-phase" excitations of the two superfluids, respectively. The energy scales for excitations in the two channels can be well separated. In fact, introducing the coupling constant $g$ as the mean between the intracomponent ones, $g_{\downarrow\downarrow}$ and $g_{\uparrow\uparrow}$, it is possible to define a spin interaction energy $\mu_s = n\,\delta g/2 $, where $\delta g = (g-g_{\downarrow\uparrow})$, in analogy with the definition of the chemical potential $\mu_d = n(g+g_{\downarrow\uparrow})/2$. As we will discuss in detail later, the presence of two degrees of freedom is associated with the emergence of two  Bogoliubov modes. 

An additional tool that enriches the physics of the mixtures is a coherent coupling provided by external radiation. This works as a ``phase locker" between the two phases associated with the two states. While the density channel is substantially unaffected, the spin sector is greatly modified, and its energy in local density approximation reads
\begin{equation}
    E = n\,\delta g  Z^2/2 -\hbar \Omega_R \sqrt{1-Z^2}\cos \phi - \hbar \delta_\mathrm{eff}Z \, ,
    \label{eq:MagnEnergy}
\end{equation}
where we introduced the normalized magnetization $Z=s/n \in [-1, +1]$, in analogy with the common two-level system. Here, $\Omega_R$ is the strength of the coupling, $\phi$ is the relative phase between the coupling field and the two-level system, while $\delta_\mathrm{eff}$ is the effective detuning from the atomic resonance. Notice that, in general, an imbalance $\Delta g = \left(g_{\downarrow\downarrow}-g_{\uparrow\uparrow}\right)/2$ between the intracomponent coupling constants contributes to $\delta_\mathrm{eff}$ as a density-dependent mean-field energy shift. This shift can always be experimentally compensated by the detuning from the atomic resonance $\delta$ (Fig.~\ref{fig:fig1}). 

At resonance, the magnetic ground state of the system can be found for a vanishing relative phase and it obeys the equation $\left( n\delta g+ \hbar \Omega_R/\sqrt{1-Z^2} \right) Z = 0$
which exhibits two different regimes. For $\hbar \Omega_R > -n\,\delta g$, the ground state is unpolarized, $Z=0$, with homogeneous magnetization all over the sample. This is always true for a miscible mixture, since $n\,\delta g >0$ \cite{Pitaevskii16}. For an immiscible mixture, when $\hbar \Omega_R < - n\,\delta g $ the ground state bifurcates into two degenerate ground states with $Z = \pm (1-\left(\hbar \Omega_R/n|\delta g|\right)^2)^{1/2}$ \cite{Recati2022}. The existence of a bifurcation in the ground state of the system is reminiscent of the quantum phase transition (QPT) occurring in the Ising model, belonging to the same universality class. For this reason, the unpolarized and bifurcated regimes are referenced as the paramagnetic and ferromagnetic phases, respectively. The existence of this bifurcation \cite{Zibold2010} and the generation of squeezed atomic states\cite{Strobel2014} was first observed with a coherently-coupled gas of Rubidium atoms in single mode approximation.

\section{The atomic system}
\label{Atomic}

\begin{figure}[t!]
 \centering
  \includegraphics[width = .35\textwidth]{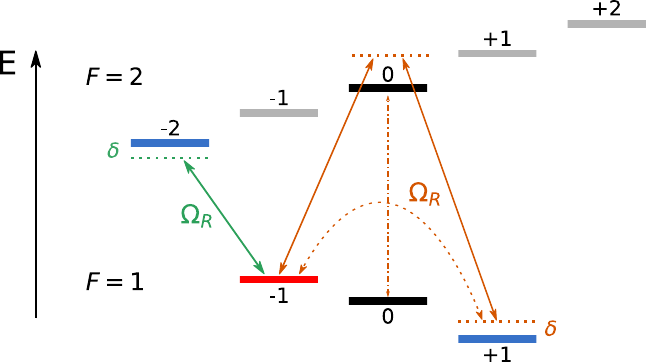}
  \caption{
  Energy levels for the $F=1,2$ hyperfine states of sodium electronic ground state. Possible coupling schemes are shown for stable two-component mixtures.
  }
   \label{fig:fig1}
\end{figure}

Spin mixtures of ultracold atoms can be realized by populating different hyperfine states $\ket{F, m_F}$ of the electronic ground state, where $F$ is the total angular momentum and $m_F$ the projection on the quantization axis. Figure~\ref{fig:fig1} shows the $F=1,2$ hyperfine manifold which is shared among most of the bosonic alkali atoms.

In the specific case of atomic sodium, the distance between the two manifolds is 1.771 GHz, to which Zeeman shift on the single state adds up in the presence of an external magnetic field. At low magnetic fields, the linear Zeeman regime dominates and the energy shift is about 0.7 MHz/G for each quantum of $m_F$. The choice of the states forming the mixture follows two requirements: the stability against spin-changing collisions and the desired combination of intra- and intercomponent interaction strengths. 

\begin{table}[b!]
    \centering
    \footnotesize
    \begin{tabular}{c|c|c|c|c|c|c|c}
         \multirow{2}*{\rotatebox[origin=c]{90}{Mix.}}&State& State& $a_{11}$ & $a_{22}$ & $a_{12}$ & $n\,\delta g/h $& $n\,\Delta g/h$\\
         & $\ket{\downarrow}$ & $\ket{\uparrow}$ & $[a_0]$&  $[a_0]$&  $[a_0]$&  [kHz] & [kHz]\\
         \hline
         \hline
         $A$&$\ket{1,-1}$ & $\ket{1,+1}$ & 54.5 & 54.5& 50.8 & 0.2  & 0\\
         $B$&$\ket{1,-1}$ & $\ket{2,-2}$ & 54.5 & 64.3& 64.3 & -1.2 & -1.2\\
    \end{tabular}
    \caption{Hyperfine components, scattering lengths, typical spin interaction energy and mean-field energy difference between state 1 and 2 for the two different spin mixtures used in the experiments. The energy values refer to the peak density at the center of the condensate. 
    }
    \label{tab:my_label}
\end{table}

In the experiments presented in the following, two spin mixtures, trapped in an elongated harmonic confinement, are used to address the cases of positive and negative $\delta g$, see Tab.~\ref{tab:my_label}. Combining $\ket{1,-1}$ and $\ket{1,+1}$ (mix. $A$), the mixture is symmetric in the interaction strengths and requires a two-photon coupling. Decay into the $\ket{1,0}$ state due to spin-changing collisions is prevented by adding a quasi-resonant microwave radiation between $\ket{1,0}$ and $\ket{2,0}$ to lift the energy of $\ket{1,0}$. The stretched state combination $\ket{1,-1}$ and $\ket{2,-2}$ (mix. $B$) is stable against spin collisions, it requires a single-photon coupling, and it is not symmetric in the intracomponent scattering lengths.
It is worth noticing that in the absence of any coupling radiation, mixture $A$ is fully miscible, i.e., both gases tend to acquire the same spatial distribution in their ground state \cite{Bienaime2016,Fava2018}, whereas mixture $B$ is immiscible and the two gases occupy different regions with minimal overlap. 

For both mixtures studied here, the energy difference between the involved states is sensitive to first-order Zeeman effect. Therefore any field fluctuation uncontrollably modifies the two-level system, causing strong shot-to-shot fluctuations in the relative population and a rapid drop of in-shot coherence between the two states.

\begin{figure*}[t!]
  \centering
  \includegraphics[width = 1.0 \textwidth]{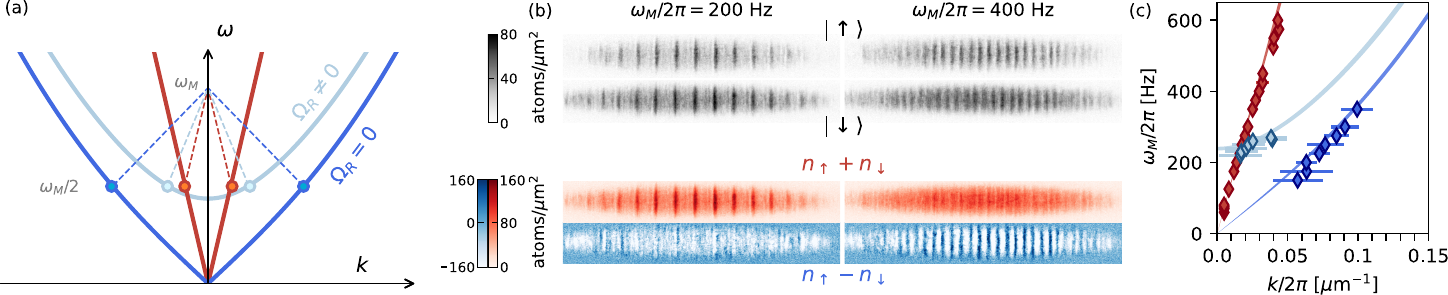}
  \caption{Elementary excitations in spin mixtures. (a) Generation mechanism of Faraday excitations in spin mixtures. An excitation with frequency $\omega_M$ decays in two phonons with half the energy and opposite wavevectors. In the absence of coupling, the density (red) and spin (blue) decay channels correspond to gapless excitations. For finite coupling, $\Omega_R \neq 0$, the spin mode (light blue) is gapped and quadratic for low wavevectors. 
  b) Two-dimensional imaging of the atoms with an axial periodic pattern in both components (top). Depending on the experimental parameters, the pattern is more evident in the total density (left images) or in the magnetization (bottom). c) Most excited wavevectors $k$, extracted from the spatial Fourier analysis of the uncoupled (blue) and coupled (light blue) mixture at different modulation frequencies. Experimental points are compared with theoretical dispersion relations (colored lines), calculated from Eq.~\ref{eq:DispD}-\ref{eq:DispS}  with no free parameters. Figures and data adapted from \cite{Cominotti2022}. }
   \label{fig:fig2}
\end{figure*}

To circumvent this problem, the field needs to be well enough stable so that the noise on the Zeeman energy can be much smaller than both the coupling strength and the spin interaction energy. Given the relatively small value of the latter, typically of the order of a few hundred Hz, see Tab.~\ref{tab:my_label}, the magnetic field has to be stable at the level of a few $\mathrm{\mu}$G.

The magnetic field stability requirements can in principle be reached with an active feedback loop for magnetic field stabilization \cite{Dedman07} or by passive means. This last solution is based on a carefully designed magnetic shield, built with a high-magnetic-permeability material such as $\mathrm{\mu}$-metal, and installed around the science chamber. This setup was already employed in previous works \cite{Esslinger98, Bloch99}, and it has the key advantage of providing low-frequency stability without the need for a magnetic field sensor. In our setup, it allows for stability at the level of a few $\mu$G  \cite{Farolfi2019}. 

\section{Elementary excitations}
\label{Faraday}
(\textit{Mix. A}) ---
In a two-component superfluid in the absence of coupling, the presence of two normal modes corresponds to two gapless excitation spectra, for the spontaneously broken U(1)$\times$U(1) symmetry of the system \cite{Pitaevskii16}, related to the independent conservation of the number of particles for each spin component. Both branches exhibit a sonic regime at low wavevector $k$, albeit with different speeds of sound $c_{d(s)}$, as they are directly determined by $\mu_{d(s)}$. The first experimental measurement revealing the existence of two well-distinguished speeds of sound was reported in Ref.~\cite{Kim2020} with a gas of sodium atoms. 

By adding a coherent coupling between the two states, the conservation law for the relative number of particles is explicitly broken.
As a consequence, the spin dispersion relation acquires a gap at $k=0$, reflecting the additional energy cost required to excite the relative phase. The full spectra of elementary excitations, captured by Bogoliubov theory \cite{Pitaevskii16}, are given by

\begin{equation}
    \hbar \omega_d = \sqrt{\frac{\hbar^2k^2}{2m}\left(\frac{\hbar^2k^2}{2m} + (g+g_{12})n\right)}
    \label{eq:DispD}
\end{equation}
\begin{equation}
    \hbar \omega_s = \sqrt{\left(\frac{\hbar^2k^2}{2m}+\hbar\Omega_R\right)\left(\frac{\hbar^2k^2}{2m} + n\,\delta g +\hbar\Omega_R \right)} \, 
    \label{eq:DispS}
\end{equation}
where it is evident that a finite value of the coupling strength $\Omega_R$ only affects the spin dispersion relation with a gap at $k=0$ equal to $\omega_{pl} = \sqrt{\Omega_R(\Omega_R + n\delta g/\hbar)}$, called \textit{plasma frequency}.

The full spectrum of elementary excitations, beyond the phononic regime reported in \cite{Kim2020}, can be measured for instance using Bragg spectroscopy, already employed for a measurement of the Bogoliubov spectrum in a single-component condensate \cite{Steinhauer2002}. A simpler approach, demonstrated in Ref.~\cite{Cominotti2022}, is to use Faraday patterns, which can be generated through a periodic 
excitation
of the non-linear interaction term by modulating the trapping potential at frequency $\omega_M$ \cite{Engels2007}. As a result, $k=0$ phonons at frequency $\omega_M$ are injected in the superfluid bulk, which subsequently decay into two entangled counter-propagating phonons, at half the driving frequency, following energy and momentum conservation. In a spin mixture, two decay channels are available [Fig.~\ref{fig:fig2}(a)]. 

To observe the phenomenon experimentally, we start from a fully polarized BEC in $\ket{\downarrow}$ which is converted into a homogeneously balanced mixture using a partial spin rotation, as detailed in \cite{Farolfi2021}. We subsequently modulate the radial confinement of the trap for a given time and then apply spin-selective imaging to measure the spatial arrangement of the atomic densities in the two states. Examples of final spin-sensitive distributions are shown in the grey scale images in the upper panel of Fig.~\ref{fig:fig2}(b) for two different modulation frequencies. The density and spin channel are then reconstructed as the sum and difference of the two density distributions. The result of this procedure is shown as the red and blue images in the lowest part of panel (b), where it is clear that a regular pattern is created in either of the two channels. For various modulation frequencies $\omega_M$, we extract the most excited wavevectors in the spatial Fourier transform of the density and spin distributions and reconstruct the dispersion relations.

A comparison between experimental points and the Bogoliubov prediction is shown in Fig.~\ref{fig:fig2}(c), for both the density (red) and the spin channels (blue, light blue). As expected, in the presence of a finite value of the coupling ($\Omega_R/2\pi = \SI{30}{Hz}$), the spin mode becomes gapped (light blue points). Remarkably, such a low coupling strength, under resonance conditions, is accessible only thanks to the high stability of the magnetic field ensured by the aforementioned magnetic shield installed around the science chamber. 

It is worth mentioning that besides elementary excitations, a spin mixture can support also localized topological excitations such as different kinds of vortices or solitons \cite{Kevrekidis2003,Hamner2011}.
One particular example of a two-component soliton is represented by the magnetic soliton \cite{Qu2016}, a localized solitary wave with an out-of-phase population immersed in a balanced miscible mixture. Such a magnetic defect was produced via phase imprinting \cite{Farolfi2020,Chai2020}, and its oscillation in a harmonic potential was observed thanks to its much longer stability, as compared to its density counterpart.

The presence of coherent coupling introduces an additional solitonic solution, known as sine-Gordon soliton, consisting of a domain wall in the relative phase \cite{Son2002}, with a $2\pi$ phase jump, whose width is fixed by the coupling strength. In one dimension, the presence of such a domain wall is expected to completely alter the dynamics of a solitonic excitation \cite{Qu2017}, leading to a dynamical change in the sign of the soliton mass. Even more appealing is the case of 2D planar geometries, where half-quantized vortices (with a $2\pi$ phase winding on one component and none on the other) can form \cite{Seo2015}. For instance, such exotic topological objects can spontaneously form as decay products of an extended domain wall in the presence of strong coherent coupling \cite{Gallemi2019}. Such domain wall introduces a confinement mechanism reminiscent of the problem of quark confinement in quantum chromodynamics \cite{Eto2018}.

\section{Para-Ferromagnetic QPT}
\label{Ferro}
(\textit{Mix. B}) ---
The mean-field-driven bifurcation in the ground state magnetization of a spatially extended system can be interpreted as a second-order quantum phase transition in the universality class of the transverse field Ising model \cite{sachdev}. In the language of magnetism, the spin interaction energy $n\,\delta g$ is associated with an \textit{easy-axis} ferromagnetic anisotropy, while the coupling strength $\Omega_R$ and the detuning from atomic resonance $\delta_\mathrm{eff}$ are associated with the \textit{transverse} and \textit{longitudinal} fields respectively \cite{Recati2022, Cominotti2023}.

\begin{figure}[t!]
 \centering
  \includegraphics[width = .49\textwidth]{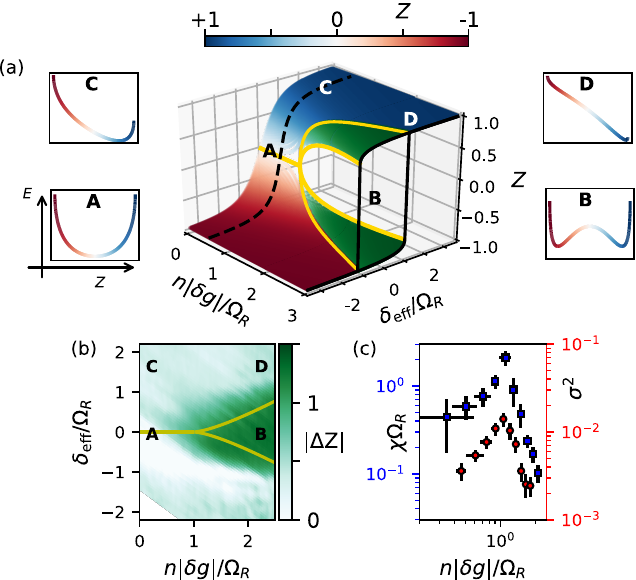}
  \caption{Ferromagnetism in mixtures. 
  (a) The central panel shows the magnetization $Z$ computed minimizing Eq. \ref{eq:MagnEnergy}, with examples shown in the side panels. The green areas 
  delimited by yellow lines mark the regions where two minima exist. A hysteresis cycle at $|n\delta g|/\hbar \Omega_R>1$ is marked by the black continuous line. The single minimum at low $|n \delta g|/\hbar \Omega_R$ and $\delta/\Omega_R=0$ (A, paramagnet) becomes a double minimum profile for $| n \delta g|/\hbar\Omega_R>1$ (B, ferromagnet). For large detuning $\delta$, only one minimum is present in both cases (C and D). (b) Measured difference in magnetization, $\Delta Z$, between systems initialized in $\ket{\downarrow}$ and $\ket{\uparrow}$, the yellow lines are the same as panel in (a). (c) The resonance of the susceptibility $\chi\Omega_R$ and of the amplitude of the magnetic fluctuations $\sigma^2$ confirm the presence of the phase transition from para- to ferromagnetism around $| n \delta g|/\hbar\Omega_R=1$. Figures and data adapted from \cite{Cominotti2023}.}
   \label{fig:fig3}
\end{figure}

The resulting phase diagram for the magnetization of the local and absolute minima of the energy from Eq.~\ref{eq:MagnEnergy} is shown in Fig.~\ref{fig:fig3}(a).
For $n|\delta g| < \hbar \Omega_R$ the system exhibits paramagnetic nature with a smooth variation of $Z$ upon a variation of $\delta_\mathrm{eff}$. In this regime, the energy landscape is single minimum, see \textbf{A} and \textbf{C} in Fig.~\ref{fig:fig3}(a). For $n|\delta g|>\hbar \Omega_R $ the system shows a hysteresis cycle in $Z_{GS}$ while varying $\delta_\mathrm{eff}$, typical of ferromagnetic materials. Within the hysteresis region (green area), the energy profile has a typical double-well structure, which is symmetric in the particular case of $\delta_\mathrm{eff} = 0$ (\textbf{B}). For finite values of detuning, the two minima have different energy until one single minimum remains for large enough $\delta_\mathrm{eff}>|\delta_\mathrm{hyst}|$(\textbf{D}), where the ferromagnet is saturated. To summarize, the phase diagram presented here contains a second-order QPT when changing $n|\delta g|/\hbar\Omega_R$ at $\delta_\mathrm{eff} = 0$, and a first-order QPT in the ferromagnetic region while varying $\delta_\mathrm{eff}$.

\begin{figure*}[t!]
 \centering
  \includegraphics[width = \textwidth]{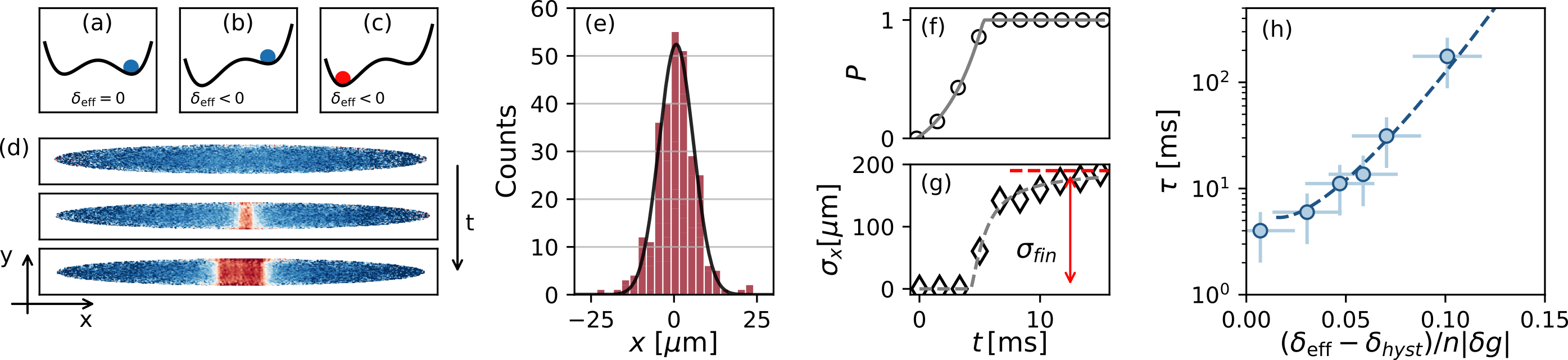}
  \caption{False vacuum decay. a) The system is prepared in the right energy well. b) By applying a non-zero detuning the system is held in the metastable state and a variable time is waited to observe the passage to the left well (c). d) Examples of experimental observation. From top to bottom: no bubble, early small bubble, late expanded bubble. e) Position of the bubbles with size $\sigma_x<\SI{50}{\mu m}$ with respect to the center of the condensate. 
  The distribution is fitted with a Gaussian function (solid black line). f) The probability $P$ to observe the bubble increases in time and saturates at one. g) The average bubble size $\sigma_x$ increases in time until it saturates to the ground state size $\sigma_\mathrm{fin}$. h) The characteristic nucleation time $\tau$ increases more than exponentially with $\delta_\mathrm{eff}$.
  Figures and data adapted from \cite{Zenesini2024}. }
   \label{fig:fig4}
\end{figure*}

A similar phase diagram was investigated in Ref.~\cite{Trenkwalder2016} with a 0-dimensional $^{39}$K BEC trapped in a spatially engineered double-well potential. In such a system, the competition between the attractive mean-field energy, which can be tuned through Feshbach resonances, and the external potential makes it possible to study the transition between a localized state on either side of the double well potential (double minimum energy landscape), to a delocalized state (single minimum). The authors used this system to characterize the transition, showing evidence of a hysteresis cycle and divergence of the susceptibility. A first investigation of the QPT with spatial extended coherently-coupled $^{87}$Rb superfluid spin mixtures was performed in Ref. \cite{Nicklas2015} quenching the strength of the coupling from the para- to the ferromagnetic side, to extract dynamical critical exponents associated with the fluctuation of the magnetization order parameter $Z$.

An in-depth investigation of the QPT was carried out by our group in Ref.~\cite{Cominotti2023}, working with mixture $B$, which features a negative value of $\delta g$, and is robust against spin-changing collisions.
Here, we demonstrated the spontaneous emergence of different magnetic phases in the superfluid sample, whose spatial extension can be controlled via the coupling strength $\Omega_R$, and through the detuning term $\delta_\mathrm{eff}$. The nature of the QPT is assessed by mapping out the magnetic ground state phase diagram by preserving the system in the stationary state of ``local" minimal energy, initializing the system in $\ket{\downarrow}$ ($\ket{\uparrow}$) and slowly increasing (decreasing) the detuning. Furthermore, the use of a harmonic trap allows us to observe the QPT in space in a single shot, thanks to the spatial variation of $n$.

In Fig.~\ref{fig:fig3}(b) we show the measurement of the full phase diagram, resulting from the difference $\Delta Z$ between the measured magnetization for increasing and decreasing detuning across zero. The ferromagnetic region emerges when the two measurements differ from zero and is displayed in green.
The presence of hysteresis in the sample confirms the emergence of a magnetic first-order QPT, characterized by a discontinuity in the magnetic order parameter $Z$. The QPT can be explored by varying the detuning (longitudinal field in the Ising model) at a fixed $n|\delta g|/\hbar \Omega_R$. On the other hand, the presence of a second-order QPT, characterized by a discontinuity in the first derivative of the order parameter, is investigated by looking at the divergence of the magnetic susceptibility $\chi=\partial Z/\partial \delta_\mathrm{eff}$ and spatial magnetic fluctuations $\sigma^2$ at $\delta_\mathrm{eff} = 0$. As shown in panel (c), both quantities peak around the critical point, as expected from the theory of QPTs. 

At last, in the region fulfilling $n|\delta g|>\hbar \Omega_R$, it is possible to deterministically create and displace magnetic domain walls by taking advantage of the buoyant character of the spin mixture, along with the inhomogeneous total density 
profile \cite{Cominotti2023}.

\section{Metastability}
\label{FV}
(\textit{Mix. B}) --- An interesting topic of study is the relaxation dynamics of ferromagnets in the hysteresis region, where, for small $\delta_\mathrm{eff}$, one of the states becomes the absolute ground state of the system while the other acquires a metastable character \cite{Lagnese2021}.

Thanks to the correspondence between field theories and a mean-field description of the condensate, the two minima can be identified as \textit{true} and \textit{false vacuum}, respectively \cite{Coleman1977}. Quantum \cite{Callan1977} or thermal \cite{Linde1983} fluctuations are expected to trigger the decay from the metastable state to the absolute minimum. 
In this case, tunneling is not a single particle process, but rather a many-body one, which, in the field theory language, corresponds to macroscopic tunneling of the field through the ferromagnetic energy barrier. This is expected to manifest through the stochastic formation of bubbles of true vacuum inside the false vacuum background\cite{Devoto2022}. 

This theory of false vacuum decay (FVD) was introduced in the context of quantum field theory \cite{Coleman1977, Callan1977} and applied to cosmological problems, but had no experimental verification so far because of the extreme energy scales and the lack of tunable parameters to investigate. The first experimental observation was recently reported by our group \cite{Zenesini2024}, achieved taking advantage from the first-order magnetic QPT naturally present in our system and from the stability of the magnetic field. The latter becomes crucial to finely tune $\delta_\mathrm{eff}$ and hence the magnetic barrier height and the energy imbalance between the two vacua. Other schemes based on coherently-coupled ultracold gases \cite{Braden2018, Fialko2015, Jenkins2024} and on spin systems \cite{Lagnese2021} have been recently proposed to simulate the decay of a relativistic metastable state. 

Ramping $\delta_\mathrm{eff}$ across the symmetric double-well condition [Fig.~\ref{fig:fig4}(a)], we initialize an elongated condensate in the metastable state [Fig.~\ref{fig:fig4}(b)] and wait for it to decay to the absolute ground state [Fig.~\ref{fig:fig4}(c)]. We let the system evolve for a tunable time and image its spatial magnetization as shown in Fig.~\ref{fig:fig4}(d). Starting from a homogeneous configuration [Fig.~\ref{fig:fig4}(d, top)], we witness the formation of a localized region of true vacuum, called bubble (middle), which subsequently expands in the sample (bottom). 

The stochastic nature of the bubble formation mechanism characterizes not only the time, but also the location at which the bubbles nucleate. In our system the process is favored in the central region of the cloud, thanks to the $Z_2$ broken symmetry, and the spatial randomness is confirmed by the Gaussian distribution of the bubble position, which has a width much larger than the initial bubble size of less than 1\,$\mu$m  [Fig.~\ref{fig:fig4}(e)]. 

After the initial tunneling process, which preserves energy, the bubble is expected to freely expand by filling the space with the energetically favorable true vacuum region. The longer the waiting time, the higher the probability of observing a bubble. Fig.~\ref{fig:fig4}(f) and (g) show the probability of bubble appearance $P$, which increases until saturation to unit value, and of the average size of the bubble $\sigma_x$ which grows in time up to $\sigma_\mathrm{fin}$. 

In \cite{Zenesini2024} the bubble nucleation rate extracted from experimental data, akin to the one shown in panel (f), has been initially compared and confirmed with numerical simulations of the coupled Gross-Pitaevskii equations. By adapting the field theory to the atomic system, good agreement has been also reported between the experimental data and the instanton solution of the false vacuum decay process: see the line in Fig.~\ref{fig:fig4}(h). Figure ~\ref{fig:fig4}(h) highlights the strong dependence of the decay time $\tau$ on the detuning $\delta_\mathrm{eff}-\delta_\mathrm{hyst}$, spanning two orders of magnitude over a detuning change of few tens of Hz, which is accessible thanks to the high magnetic field stability in our experiment. 

The bubble formation mechanism in flattened two-dimensional systems, their collisional properties, 
meson confinement \cite{Birnkammer2022}, and the role of the observer in bubble formation \cite{Maki2023} are interesting research directions.

\section{Far-from-equilibrium dynamics}
\label{QT}
(\textit{Mix. A}) --- Spatially elongated magnetic systems with nonuniform properties allow for the study of processes arising from the coexistence of different dynamical regimes within the same sample. The spin $\vec{S}$ can be locally represented on the Bloch sphere via both the magnetization $Z$ and the phase $\phi$, and its dynamical evolution, even in a far from equilibrium regime, is exactly captured by the Landau-Lifshitz equation. In a coupled miscible superfluid mixture, this equation takes the form: $ 
    \partial_t \vec{S} + \partial_x \vec{j}_{s} = -H(\vec{S})\times \vec{S}$,
where $H=\Omega_R \hat{e}_1 + \delta g\,S_z \hat{e}_3$ is an effective non-linear magnetic field, $\partial_x \vec{j}_{s}$ is a position-dependent term related to the spin current $\vec{j}_{s}$ inside the system, and $\hat{e}_1, \hat{e}_3$ label the axes on the Bloch sphere.

Simple dynamics is expected when one of the two terms in $H$ dominates, either Rabi oscillation driven by $\Omega_R$, or the nonlinear effect coming from the anisotropy $\delta g\,S_z$. Thanks to the high stability of the coupling at low Rabi frequency, it is possible to engineer the equivalent of a magnetic material with different dynamical regimes \cite{Farolfi2021QT}. 

\begin{figure}[t!]
  \centering
  \includegraphics[width = .4\textwidth]{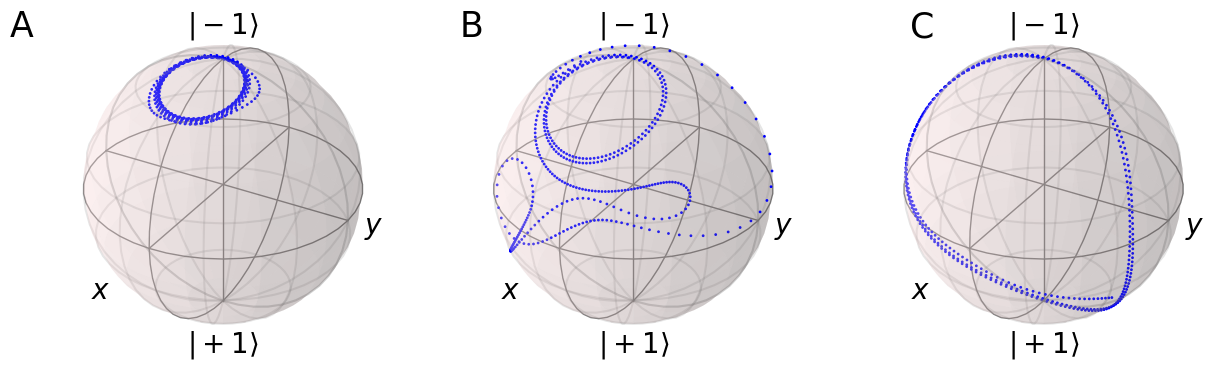}
  \includegraphics[width = .45\textwidth]{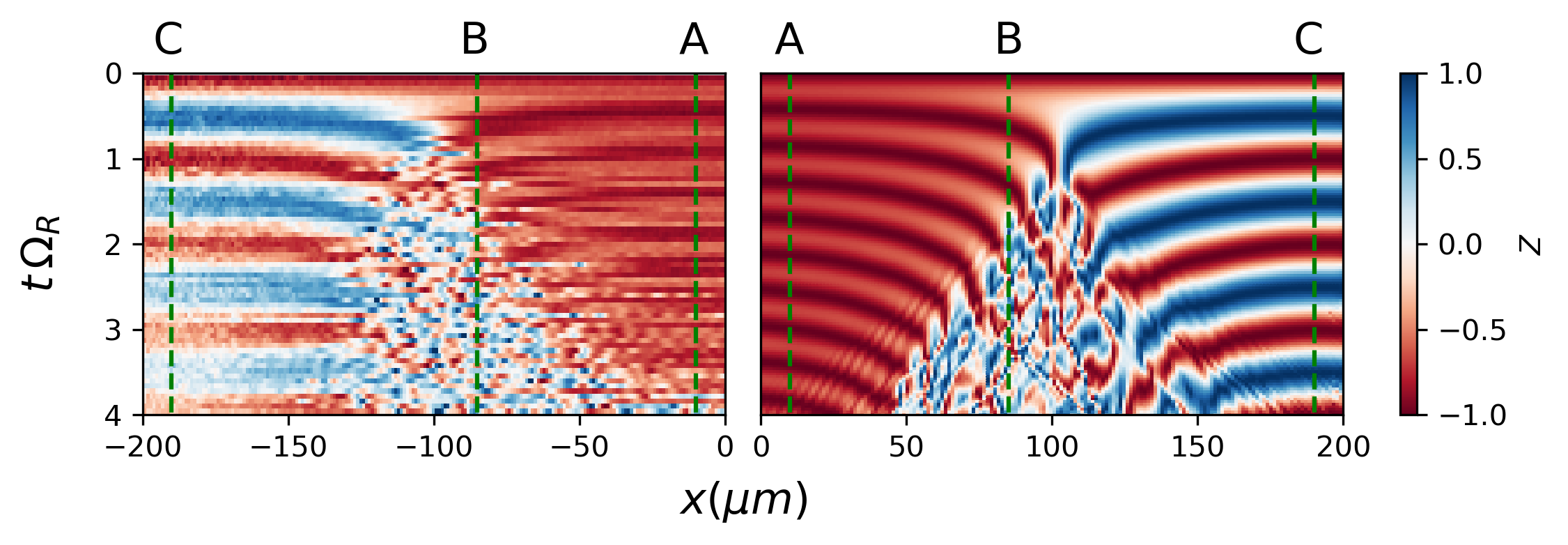}
  \caption{Dynamics in excited state. In \textbf{A} the strong non-linearity in the center of the cloud locks the motion in the vicinity of the north pole of the Bloch sphere. In the low-density tail of the cloud \textbf{C}, the motion is between the north and the south pole. Close to the interface region \textbf{B}, the motion follows simple dynamics just for a short time, being then affected by the quantum torque which drives a chaotic motion. Looking at the whole cloud, this third behavior shows up as two cones of strongly fluctuating magnetization which expand and ``destroy" the inner and outer regions. The left panel contains experimental results while the right one the numerical simulations from which the Bloch spheres in A, B, C are obtained.
  Figure adapted from \cite{Farolfi2021QT}. }
   \label{fig:fig5}
\end{figure}

In the upper panels of Fig.~\ref{fig:fig5}, we show the evolution of the magnetization in time, for spins initially aligned along $z$ and suddenly subject to an external coupling field at $t=0$. In the center of the cloud, the spin precesses around a fixed point close to the north pole of the Bloch sphere (panel A), while in the low-density outer regions the spin motion encompasses the whole Bloch sphere, as expected from single-atom dynamics (panel C). Interesting dynamics occurs at the spatial interface $x_c$ between the two regions  (panel B), identified by $n(x_c)\delta g  = \hbar \Omega_R$, where both in experiments and simulations (lower panel in Fig.~\ref{fig:fig5}) the interface breaks down into a turbulent magnetization region, which expands in the sample. 
The origin of the turbulent regime is rooted in the build-up of spin currents close to the interface, where the phase is rotating in the outer side and locked in the inner one.
As a result, the gradient of the current $\partial_x j_s(x)$ gains importance, as compared to $H$, until the local orbits on the Bloch sphere deviate from the original one in the bulk (see B). The most energetic term, in this case, is proportional to $\nabla^2 \vec{S}$, is known in the literature as \textit{quantum torque} and arises from the exchange interactions \cite{Farolfi2021QT}. It is worth noticing that the highly turbulent regime appears because the system is excited and driven out-of-equilibrium from the sudden switch-on of the coupling. While a spatial magnetization gradient was also present in the work of Ref.~\cite{Cominotti2023}, there the focus was on linear excitations on the spin ground state background, and the quantum torque played a minor role.

\section{Future directions}

The studies presented here provide a brief overview of experiments carried out with spin mixtures of ultracold atoms in the presence of coherent coupling. The Hamiltonian for the spin sector of the system shares the same universality class of the quantum Ising model, allowing for the study of magnetic phenomena and excitations above the ground state in a controlled yet flexible environment, where a paradigmatic example is the experimental observation of false vacuum decay. A natural extension of this topic is the study of the effect of temperature, which is expected to exponentially suppress the decay rate until quantum effects set in \cite{Linde1983, Ng2021}. Another promising possibility is the implementation of a dedicated hardware, consisting of a pair of localized Raman coupling laser beams \cite{Zou2021} to generate bubbles on demand, possibly in a 2D geometry, opening to the study of the expansion dynamics of bubbles.

The advanced spatial engineering of both the coupling and the density profiles are powerful tools to design atomic and spintronic circuits or to create vortices on demand in novel geometric configurations. Another promising direction is the extension of the studies of fragmentation, as in 0D sodium condensates \cite{Evrard21, JimenezGarcia2019}, to higher dimensions, by using spatially-extended spinor gases in magnetic fields-free environments, which are becoming available using magnetic shields \cite{Rogora2024}.

\section{Acknowledgements}
We thank all the members of the Pitaevskii BEC Center in Trento for stimulating discussions.
We acknowledge funding from Provincia Autonoma di Trento, from the European Union’s Horizon 2020 research and innovation Programme through the STAQS project of QuantERA II (Grant Agreement No. 101017733), from the European Research Council (ERC) under the European Union’s Horizon 2020 research and innovation program (grant agreement No 804305),  and from PNRR MUR project PE0000023-NQSTI. 
This work was supported by Q@TN, the joint lab between University of Trento, Fondazione Bruno Kessler, National Institute for Nuclear Physics and National Research Council.



%

\end{document}